\documentclass[aps,pra,superscriptaddress,amsmath,amssymb,preprintnumbers]{revtex4}

\usepackage{subfigure}
\usepackage{color}
\usepackage{soul}
\usepackage{amsmath}
\usepackage{amssymb}
\usepackage{booktabs}
\usepackage{graphicx}   
\usepackage{bm}
\usepackage{color}
\usepackage{cancel}
\usepackage{slashed}
\usepackage{subfigure}
\usepackage{soul}

\newcommand{\Ignore}[1]{}

\newcommand{\Ket}[1]{\left\vert #1\right\rangle}
\newcommand{\Bra}[1]{\left\langle #1\right\vert}

\newcommand{\BraKet}[2]{\left\langle#1\vert #2\right\rangle}

\newcommand{\ii}{\mathrm{i}}
\newcommand{\ee}{\mathrm{e}}
\newcommand{\idop}{1\!\!1}
\newcommand{\diff}{\mathrm{d}}

\newcommand{\hplus}{H_+}
\newcommand{\hminus}{H_-}

\newcommand{\variance}{\Delta}

\begin{document}


\pagecolor{white}

\title{Short-time behavior of a system ruled by non-Hermitian time-dependent Hamiltonians}

\author{Benedetto Militello}
\author{Anna Napoli}

\affiliation{Universit\`a degli Studi di Palermo, Dipartimento di Fisica e Chimica - Emilio Segr\`e, Via Archirafi 36, 90123 Palermo, Italia}
\affiliation{I.N.F.N. Sezione di Catania, Via Santa Sofia 64, I-95123 Catania, Italia}

\begin{abstract}
The short-time behavior of the survival probability of a system governed by a time-dependent non-Hermitian Hamiltonian is derived using to the second order perturbative approach. 
The resulting expression allows for the analysis of some situations which could be of interest in the field of quantum technology. For example, it becomes possible to predict a quantum Zeno effect even in the presence of decay processes. 
\end{abstract}

\maketitle


\section{Introduction}\label{sec:Introduction}

In the standard formulation of quantum mechanics, a closed and isolated quantum system is assumed to be governed by a time-independent Hermitian Hamiltonian. Hermiticity ensures both a real spectrum of the Hamiltonian and unitarity of the time evolution, whose physical meanings are, respectively, the meaningfulness of the values of the energy and the preservation of probability~\cite{ref:TB-Sakurai,ref:TB-Messiah}. Time independence, on the other hand, allows to evaluate the time evolution in a reasonably easy way, through simple exponentiation of the Hamiltonian operator.
Generally speaking, when studying a system interacting with \lq something else\lq\,  (a driving classical fields, a reservoir, etc.), provided all the degrees of freedom are included, one can always obtain a description through a time-independent Hermitian Hamiltonian. However, when only the degrees of freedom of the system are taken into account, time-dependence, non-Hermiticity or both will arise in the relevant effective description.
For example, the action of a classical time-dependent field on a quantum system is responsible for the Hamiltonian becoming time-dependent, which in general makes it harder to solve the dynamical problem. It is worth noting that the time-independent (Hermitian) counterpart involves the free Hamiltonians of both the system and the electromagnetic field, their interaction terms, as well as the field in a classical state, usually a coherent state~\cite{ref:TB-MandelWolf}.
Moreover, the presence of an interaction with an environment in a thermal state, under suitable hypotheses, can be analyzed through approximate methods, such as master equations~\cite{ref:TB-Gardiner,ref:TB-BreuerPetruccione} and, in some cases, through non-Hermitian Hamiltonians. The exploitation of non-Hermitian Hamiltonians is traceable back to Gamow's phenomenological model for the $\alpha$-decay~\cite{ref:Gamow}, where the imaginary part of the eigenvalues of the Hamiltonian represents the decay rates of the eigenstates, which are not stationary states as in the Hermitian case. The Feshbach-Porter-Weisskopf theory for neutron scattering is also based on a model involving a complex potential whose imaginary part describes an absorption process~\cite{ref:FeshbachPorterWeisskopf}.

Time-dependent Hamiltonians are commonly used, whereas non-Hermitian physics is generally considered weird, although it arises in several physical contexts, ranging from acoustics~\cite{ref:Gu2021} to optics~\cite{ref:Ruter2010,ref:Cao2020} to quantum systems, such as superconducting devices~\cite{ref:Cayao2022,ref:Ohnmacht2024,ref:Weijian2021}, waveguides~\cite{ref:Tebbenjohanns2024} and optomechanical systems~\cite{ref:Pino2022,ref:Sun2023,ref:Martinez-Azcona2025}. We also mention pseudo-Hamiltonian operators, which are nothing but coefficient matrices involved in specific linear equations in mean field theories, often involving both gain and loss processes, leading to eigenvalues with positive and negative imaginary parts \cite{ref:ShuoLiu2020,ref:Ming-Jie2024,ref:XizhouShen2024}. 
Last but not least, time-dependent non-Hermitian Hamiltonians have been used to describe several lossy systems driven by external fields, such as atomic systems with unstable states subjected to Stimulated Raman Adiabatic Passage~\cite{ref:Vitanov1997,ref:Scala2010} or Landau-Zener processes in lossy systems~\cite{ref:Militello2019,ref:Militello2020-LZ}.
As a consequence of the growing interest in non-Hermitian physics, several theoretical studies have been reported over the decades. In fact, exactly solvable non-Hermitian models have been presented~\cite{ref:Fring2021} and general techniques based on a perturbative approach have been proposed and exploited focusing on specific situations~\cite{ref:Sternheim1972,ref:Buth2004,ref:Longhi2017}, or studying general dynamical features~\cite{ref:Hajong2024}. 
Statistical and thermodynamic properties in non-Hermitian physics have also been investigated~\cite{Sergi2015,ref:Carlstrom2020,ref:Zhou2021,ref:Cipolloni2023,ref:Turkeshi2023,ref:Erdamar2024}. 
Among the non-Hermitian operators, some classes deserve particular attention. First, the {\cal PT}-symmetric operators~\cite{ref:Bender1998,ref:Bender2003}, which are characterized by the possibility of having real spectra and inducing unitary evolutions. Another interesting type of non-Hermitian operators is given by the class of normal operators~\cite{ref:Macklin1984}, identified by the property of commuting with their adjoint operators, implying that their left and right eigenstates are adjoint to each other, while the eigenvalues are complex numbers. When the non-Hermiticity arises as an effective description derived from a Lindblad master equation, then the non-Hermitian part of the Hamiltonian turns out to be a non-negative operator. Indeed, in such a case, the non-Hermitian part of the Hamiltonian essentially emerges as the operator involved in the anticommutator of the master equation with a non negative operator~\cite{ref:Militello2020}.

In this paper we focus on the short-time behavior of the survival probability (i.e., the probability of finding the system in its initial state) of general time-dependent non-Hermitian systems. In fact, although the survival probability has been studied in systems which are non-Hermitian~\cite{ref:Yamada2023} or time-dependent~\cite{ref:Majeed2021,ref:Suzuki2024}, there is limited literature specifically concerning the survival probability after a short-time evolution when the Hamiltonian is both time-dependent and non-Hermitian. This topic could be of interest in view of possible applications to the quantum Zeno effect, which occurs in the short-time dynamics, usually evaluated with a second-order perturbative approach as in the original formulation by Misra and Sudarshan~\cite{ref:MisraSudarshan}. Also, as we will see, our results concretely predict the possibility of effectively suppressing the effects of decays and general non-Hermitian processes, when the Hamiltonian has a specific time dependence. Moreover, the closed expression for the survival probability we report here turns out to be very helpful also in the Hermitian case, providing an intriguing generalization of the well-known result due to Misra and Sudarshan in the study of the quantum Zeno effect.
It is worth emphasizing that by exploiting second-order perturbation theory, we have derived a closed-form expression for the survival probability that applies to a general Hamiltonian --- whether Hermitian or non-Hermitian, time-dependent or time-independent --- without specifying the physical system. The resulting formula therefore has the advantage of being applicable to a wide range of scenarios, in contrast to very recent works where the survival probability is computed either for arbitrary driven open quantum systems~\cite{ref:Majeed2021}, or for specific models such as a time-dependent Ising chain coupled to a bosonic bath at zero temperature~\cite{ref:Suzuki2024}.
Nevertheless, the applicability of a second-order perturbative treatment deserves specific considerations related to the specific models that we will analyze in the section dedicated to the applications. 
The paper is organized as follows. In Sec.~\ref{sec:Results} we show the main result of this paper, which consists of a closed expression for the short-time survival probability, while in Sec.~\ref{sec:Methods} it is shown how to reach this result. In Sec.~\ref{sec:Applications} some applications are presented, ranging from the quantum Zeno effect in Hermitian systems to the decay suppression for unstable systems. Finally, in Sec.~\ref{sec:Discussion} we provide a discussion on the general results and relevant applications. An extensive discussion about the validity of the approximate method we use is also provided in this section.


\section{General results}\label{sec:Results}

Considered a system governed by a generic Hamiltonian, which could be time-dependent and non-Hermitian, we present here the second-order survival probability derived through the Dyson expansion of the relevant Schr\"odinger equation. Subsequently, we specialize this formula to some particular cases which will be of interest in connection with the applications.

\subsection{Second Order Survival Probability}\label{sec:SecondOrderSurvival}

The second-order survival probability for a system governed by a time-dependent non-Hermitian Hamiltonian can be written in the following form (see Sec.~\ref{sec:Methods} for the derivation):
\begin{eqnarray}\label{eq:2OrdSurvProb-basic}
\nonumber
&& |\Bra{\psi} {\cal T}(t_2, t_1) \Ket{\psi} |^2_{(2)} =  
1 - 2 \,\ii \Bra{\psi} \int_{t_1}^{t_2} \hminus(s)\diff s \Ket{\psi}  - \variance_+^2 \left( \int_{t_1}^{t_2} \hminus(s) \diff s, \Ket{\psi} \right) \\
\nonumber
&& \qquad - \variance_-^2 \left( \int_{t_1}^{t_2} \hplus(s) \diff s, \Ket{\psi} \right)  \\
&& \qquad - \int_{t_1}^{t_2} \diff s \int_{t_1}^{s} \diff s' \Bra{\psi} \left( [\hplus(s), \hminus(s')] - [\hplus(s'), \hminus(s)] \right)  \Ket{\psi} \,,
\end{eqnarray}
where the subscript $(2)$ indicates quantities evaluated up to the second order in the perturbation treatment, the operators 
\begin{eqnarray}\label{eq:HpHm}
H_\pm(t) = \frac{H(t) \pm H^\dag(t)}{2} \,,
\end{eqnarray}
are the Hermitian ($\hplus$) and anti-Hermitian ($\hminus$) parts of the Hamiltonian operator, and
\begin{eqnarray}\label{eq:Variances}
\variance_\pm^2 (A, \Ket{\psi}) = \Bra{\psi}A^2 \Ket{\psi} \pm (\Bra{\psi}A\Ket{\psi})^2 \,.
\end{eqnarray}
It is the case to observe that the functional $\variance_-^2$ coincides with the ordinary variance of the operator $A$ related to the state $\Ket{\psi}$.
The last term, which we will address as the cross term, is the double-time integrated difference between the commutators of the Hermitian and anti-Hermitian parts of the Hamiltonian at different times, with inverse orders. A direct physical interpretation of this term is not straightforward, although it clearly encodes the non-trivial interplay between the coherent and dissipative dynamics of the system. The presence of commutators in the last term of eq. \eqref{eq:2OrdSurvProb-basic} between the Hermitian and anti-Hermitian parts of the Hamiltonian at different times reflects, in a sense, how the time ordering of these two contributions affects the evolution. In particular, the difference between two commutators evaluated at inverted times points to the emergence of time-asymmetric effects due to the non-Hermitian nature of the Hamiltonian. This is a natural feature in the context of open quantum systems.

It is well known that the Dyson expansion is not equivalent to a Taylor expansion in time. Indeed, while for time-independent Hamiltonians the $k$-th order Dyson term (the $0^{th}$ order term being the identity) is proportional to $t^k$, in general, for time-dependent Hamiltonians, contributions proportional to $t^k$ can be present in all Dyson terms from the $1^{st}$ to the $k^{th}$.
Nevertheless, because of the specific form of the second order correction reported in \eqref{eq:2OrdSurvProb-basic}, when one wants to make the Taylor series truncated to the second order, it turns out that the cross term of the survival probability does not contribute. This is a consequence of the fact that they involve a double integration with respect to time and that the zeroth order term of the integrand is vanishing. Indeed, as discussed in the next subsection, time-independent Hamiltonians give rise to a vanishing contribution from the cross terms.

\subsection{Special cases}\label{sec:SpecialCases}

It is interesting to observe that there are some situations in which the cross term of \eqref{eq:2OrdSurvProb-basic} gives zero contribution, implying the survival probability to assume the following form:
\begin{eqnarray}\label{eq:2OrdSurvProb-normal}
\nonumber
&& |\Bra{\psi} {\cal T}(t_2, t_1) \Ket{\psi} |^2_{(2)} =  
1 - 2 \,\ii \Bra{\psi} \int_{t_1}^{t_2} \hminus(s) \diff s \Ket{\psi} - \variance_+^2 \left( \int_{t_1}^{t_2} \hminus(s) \diff s, \Ket{\psi} \right) \\
&& \qquad - \variance_-^2 \left( \int_{t_1}^{t_2} \hplus(s) \diff s, \Ket{\psi} \right)  \,,
\end{eqnarray}
which means that the Hermitian and anti-Hermitian components act on the system independently, when one focuses on the (second order) survival probability. 

Two sufficient conditions under which the cross term vanishes are the following: 
\begin{eqnarray}
&& [ \hplus(t), \hminus(t') ]  = 0 \,, \qquad \forall t, t'\,, \\
&& \partial_t\hplus = \partial_t\hminus = 0 \,.
\end{eqnarray}

In the first case, both commutators are identically zero.
In the second case, i.e., for time-independent Hamiltonians, it is straightforward to see that the two commutators are equal so that their difference is zero: $[\hplus(s), \hminus(s')]$ $- [\hplus(s'), \hminus(s)]$ $= [\hplus, \hminus]$ $- [\hplus, \hminus] = 0$\,.

In the very special case of a Hermitian Hamiltonian, the second-order approximated survival probability can be expressed as $1$ diminished by the variance of the operator obtained integrating the Hamiltonian in the relevant time interval:
\begin{equation}\label{eq:SurvProb_Variance}
|\Bra{\psi} {\cal T}(t_2, t_1) \Ket{\psi} |^2_{(2)} = 1 - \variance_-^2\left( \int_{t_1}^{t_2} \hplus(s) \diff s, \Ket{\psi} \right) \,.
\end{equation}
This is the natural, though not obvious, extension of the result for time independent Hamiltonian, $|\BraKet{\psi}{\psi(t)}|^2 \approx 1-\variance_-^2(H, \Ket{\psi}) t^2$, which is the ground for the analysis of the quantum Zeno effect for time-independent Hamiltonians.


\section{Methods}\label{sec:Methods}

Let us now prove \eqref{eq:2OrdSurvProb-basic}. To this end, consider that the short-time dynamics can be evaluated through the Dyson series truncated to the second order, even in the case of non-Hermitian Hamiltonian:
\begin{eqnarray}
{\cal T}(t_2, t_1)_{(2)} =  \idop - \ii \int_{t_1}^{t_2} H(s) \diff s - \int_{t_1}^{t_2} \diff s \int_{t_1}^{s} \diff s' H(s) H(s') \,,
\end{eqnarray}
and
\begin{eqnarray}
{\cal T}(t_2, t_1)_{(2)}^\dag =  \idop + \ii \int_{t_1}^{t_2} H^\dag(s) \diff s - \int_{t_1}^{t_2} \diff s \int_{t_1}^{s} \diff s' H^\dag(s') H^\dag(s) \,.
\end{eqnarray}

From these we get, keeping only terms up to the second order:
\begin{eqnarray}\label{eq:ProofChainNH2}
\nonumber
&& |\Bra{\psi} {\cal T}(t_2, t_1) \Ket{\psi} |^2_{(2)}  = \Bra{\psi} {\cal T}(t_2, t_1)_{(2)} \Ket{\psi} \Bra{\psi} {\cal T}(t_2, t_1)_{(2)}^\dag \Ket{\psi} =  \\
\nonumber 
&& \quad 1  - \ii \Bra{\psi} \int_{t_1}^{t_2} H(s) \diff s \Ket{\psi} 
+ \ii \Bra{\psi} \int_{t_1}^{t_2} H^\dag(s) \diff s \Ket{\psi} \\
\nonumber
&& \quad + \left( \Bra{\psi} \int_{t_1}^{t_2} H(s) \diff s \Ket{\psi} \right) \left( \Bra{\psi} \int_{t_1}^{t_2} H^\dag(s') \diff s' \Ket{\psi} \right) \\
&& \quad - \Bra{\psi} \int_{t_1}^{t_2} \diff s \int_{t_1}^{s} \diff s'  \left( H(s) H(s') + H^\dag(s') H^\dag(s) \right) \Ket{\psi}  \,.
\end{eqnarray}
After explicitly writing down $H(t) = \hplus(t) + \hminus(t)$ and $H^\dag(t) = \hplus(t) - \hminus(t)$, one can separate the contributions according to the following analysis.

First, the two first-order terms in the first line of \eqref{eq:ProofChainNH2} yield the term proportional to the integral of $\hminus$.

Second, the product of the time integrals of the average values of  $H$ and $H^\dag$ in the third line of \eqref{eq:ProofChainNH2} gives:
\begin{eqnarray}
&& \nonumber
\left( \Bra{\psi} \int_{t_1}^{t_2}  (\hplus(s) + \hminus(s)) \diff s \Ket{\psi} \right) \left( \Bra{\psi} \int_{t_1}^{t_2} (\hplus(s')-\hminus(s))  \diff s' \Ket{\psi} \right) = 
\\
&& \quad \left(\Bra{\psi} \int_{t_1}^{t_2}  \hplus \Ket{\psi}\right)^2 - \left( \Bra{\psi} \int_{t_1}^{t_2}  \hminus \Ket{\psi}\right)^2 \,,
\end{eqnarray}
which are involved in the $\variance$'s.

Let us now focus on the integrand operator of the fourth line:
\begin{eqnarray}
\nonumber
&& H(s) H(s')  + H^\dag(s') H^\dag(s) = \hplus(s) \hplus(s') + \hplus(s) \hminus(s') \\
\nonumber
&& \quad \qquad + \hminus(s) \hplus(s') + \hminus(s) \hminus(s') + \hplus(s') \hplus(s) \\
&&  \quad \qquad - \hplus(s') \hminus(s) - \hminus(s') \hplus(s) + \hminus(s') \hminus(s) \,.
\end{eqnarray}
Since $\int_a^b \diff s \int_a^s \diff s' f(s,s') = \int_a^b \diff s' \int_{s'}^b \diff s f(s,s')$ (they are two equivalent ways to span the points of the same triangular dominion), we have:
\begin{eqnarray}
\nonumber
&&   \int_{t_1}^{t_2} \diff s \int_{t_1}^{s} \diff s' ( \hplus(s) \hplus(s') + \hplus(s') \hplus(s) ) = \\
\nonumber
&& \qquad  \int_{t_1}^{t_2} \diff s \int_{t_1}^{s} \diff s'  \hplus(s) \hplus(s')  + 
 \int_{t_1}^{t_2} \diff s' \int_{s'}^{t_2} \diff s \hplus(s') \hplus(s)  = \\
\nonumber
&& \qquad   \int_{t_1}^{t_2} \diff s \int_{t_1}^{s} \diff s' \hplus(s) \hplus(s')  + 
 \int_{t_1}^{t_2} \diff s \int_{s}^{t_2} \diff s' \hplus(s) \hplus(s')  = \\
&& \qquad  \int_{t_1}^{t_2} \diff s \int_{t_1}^{t_2} \diff s' \hplus(s) \hplus(s') = 
\int_{t_1}^{t_2} \diff s \, \hplus(s) \int_{t_1}^{t_2} \diff s' \hplus(s') = \left(  \int_{t_1}^{t_2} \diff s \, \hplus(s) \right)^2 \,,
\end{eqnarray}
where, in passing from the second to the third line the variables $s$ and $s'$ have been swapped ($s \rightleftarrows s'$).
Similarly,
\begin{eqnarray}
\nonumber
 \int_{t_1}^{t_2} \diff s \int_{t_1}^{s} \diff s' ( \hminus(s) \hminus(s') + \hminus(s') \hminus(s) ) = 
 \left( \int_{t_1}^{t_2} \diff s \, \hminus(s) \right)^2 \,.
\end{eqnarray}
The expectation values of these operators are involved in the $\variance$'s. Finally, the four remaining terms (i.e., those involving products of $\hplus$ and $\hminus$) give rise to the difference of commutators in \eqref{eq:2OrdSurvProb-basic}.

Summing up all the contributions, one gets:
\begin{eqnarray}\label{eq:ProofChainNH3}
\nonumber
&& |\Bra{\psi} {\cal T}(t_2, t_1) \Ket{\psi} |^2_{(2)}  =  1  - 2 \ii \Bra{\psi} \int_{t_1}^{t_2} \hminus(s) \diff s \Ket{\psi}  \\
\nonumber
&& \quad + \left(\Bra{\psi} \int_{t_1}^{t_2}  \hplus \Ket{\psi}\right)^2 - \left( \Bra{\psi} \int_{t_1}^{t_2}  \hminus \Ket{\psi}\right)^2 \\
\nonumber
&& \quad + \left(  \int_{t_1}^{t_2} \diff s \, \hplus(s) \right)^2 +  \left(  \int_{t_1}^{t_2} \diff s \, \hminus(s) \right)^2  \\
&& \quad - \int_{t_1}^{t_2} \diff s \int_{t_1}^{s} \diff s' \Bra{\psi} \left( \hplus(s) \hminus(s') - \hminus(s') \hplus(s)  - \hplus(s') \hminus(s) + \hminus(s) \hplus(s') \right)  \Ket{\psi} \,, \,\,\,\,\,\,\,
\end{eqnarray}
which, taking into account definitions in \eqref{eq:Variances}, gives exactly \eqref{eq:2OrdSurvProb-basic}.


\section{Applications}\label{sec:Applications}\label{sec:applications}

We now focus on some applications. First, we consider the case of a time-dependent Hermitian Hamiltonian, showing the connection between the time dependence of the Hamiltonian and the occurrence of a behavior similar to the quantum Zeno effect. Second, we consider non-Hermitian Hamiltonians which describe either the state of a quantum system undergoing some decays or a set of quantities in a gain-loss scheme. The analysis of the short-time survival probability allows to bring to light the possibility to alter and even suppress the relevant non-Hermitian processes.

{\it Hermitian case ---}  Let us consider a time-dependent Hermitian Hamiltonian, so that the survival probability can be expressed as in \eqref{eq:SurvProb_Variance}.
If the system Hamiltonian has a time dependence such that its average in a time interval $[t_1,t_2]$ (which from now on will be $[0, t]$) is equal to zero then the second-order survival probability is equal to unity at that time, regardless of the initial state. For example, any oscillating time-dependence will produce this result.

It is interesting to observe that a similar result can be obtained from the Magnus expansion~\cite{ref:Magnus} in some very special cases. Indeed, such expansion allows to express the unitary evolution operator as the exponential of an operator, ${\cal T}(t) = \exp(\sum_{k=1}^{\infty}\Omega_k(t))$, with $\Omega_1=\int_0^t -\ii \, H(s) \, \diff s$, and all the other $\Omega_k$'s ($k\ge 2$) which depend on the commutators of the Hamiltonian operators at different times, $[H(s), H(s')]$. Therefore, under the assumption that the Hamiltonians at different times commute, one finds that the evolution operator is simply the exponential of the integral of the Hamiltonian, which reduces to the identity operator when such integral vanishes, giving rise to a survival probability equal to $1$. This result is exact and holds order by order in the expansion, but it applies to a special class of Hamiltonians that satisfy both conditions $\int_0^t H(s) \diff s=0$ for some $t$ and $[H(s), H(s')]=0$ for all $s,s'$. On the contrary, our analysis, though valid in the context of a second-order approximation, requires only the first condition, $\int_0^t H(s) \diff s=0$ for some $t$.

As a specific example, let us consider a physical system whose Hamiltonian is:
\begin{eqnarray}\label{eq:Hermitian-1}
H(t) = \lambda \eta \sin(\omega t) \sigma_x + \lambda (1-\eta) \cos(\omega t) \sigma_y \,,
\end{eqnarray}
with $\eta$ a dimensionless parameter and $\sigma_k$'s the Pauli matrices. 

\begin{figure}[h]
\centering
\begin{tabular}{ccc}
\subfigure[]{\includegraphics[width=0.30\textwidth, angle=0]{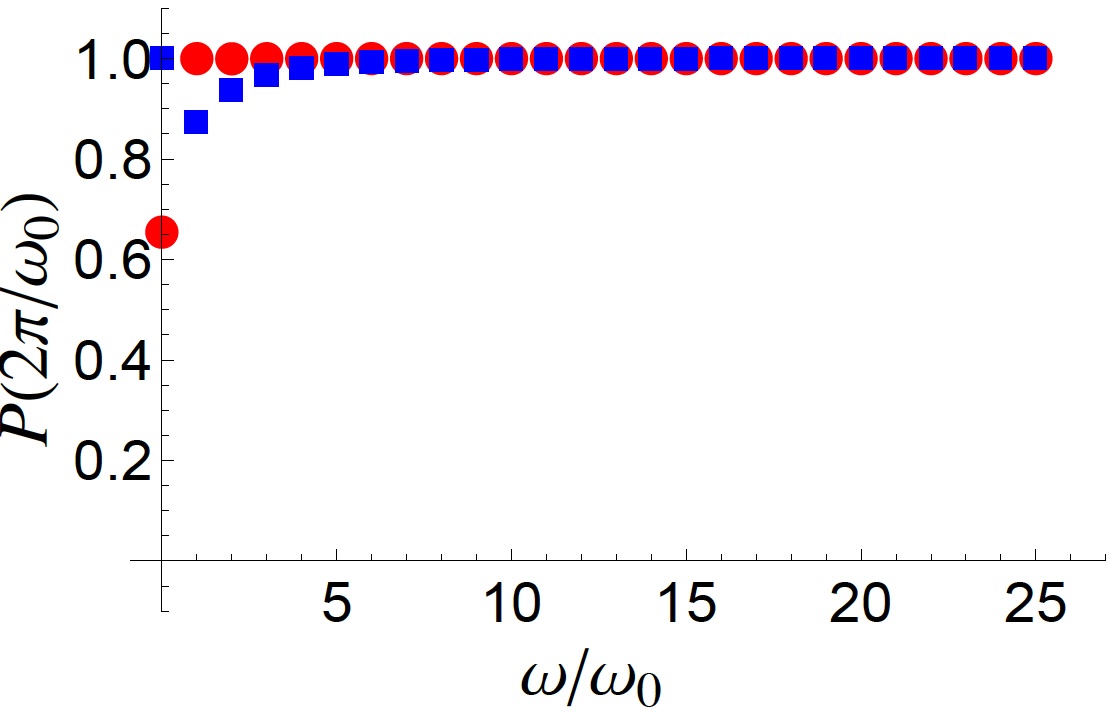}} &
\subfigure[]{\includegraphics[width=0.30\textwidth, angle=0]{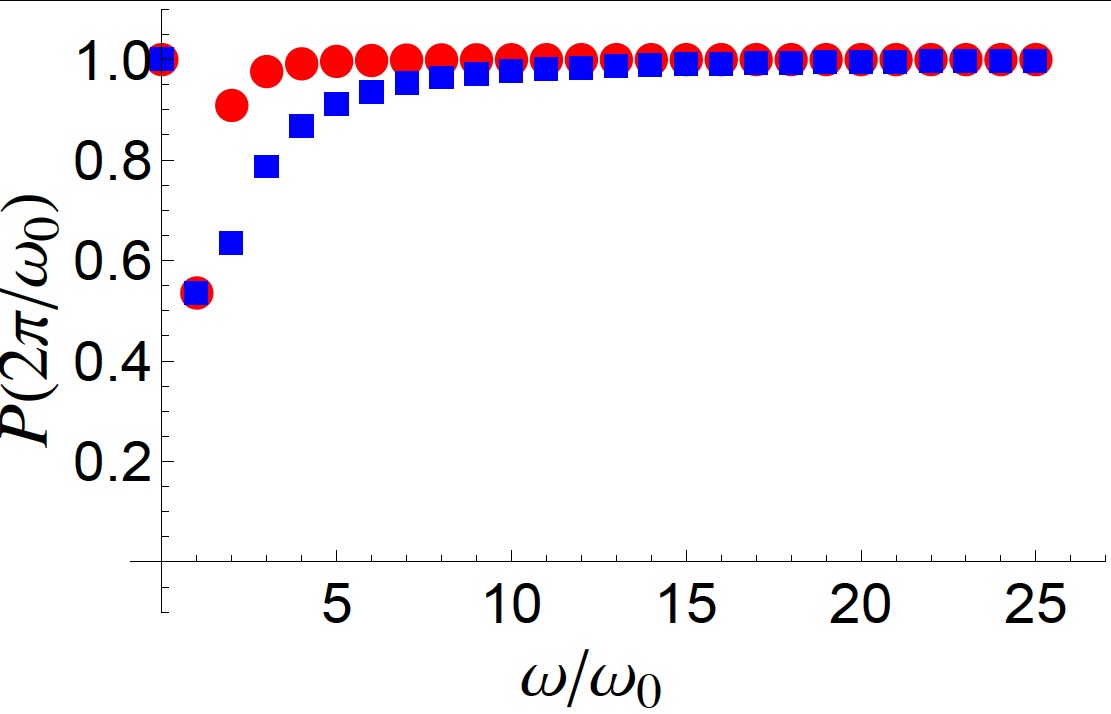}} &
\subfigure[]{\includegraphics[width=0.30\textwidth, angle=0]{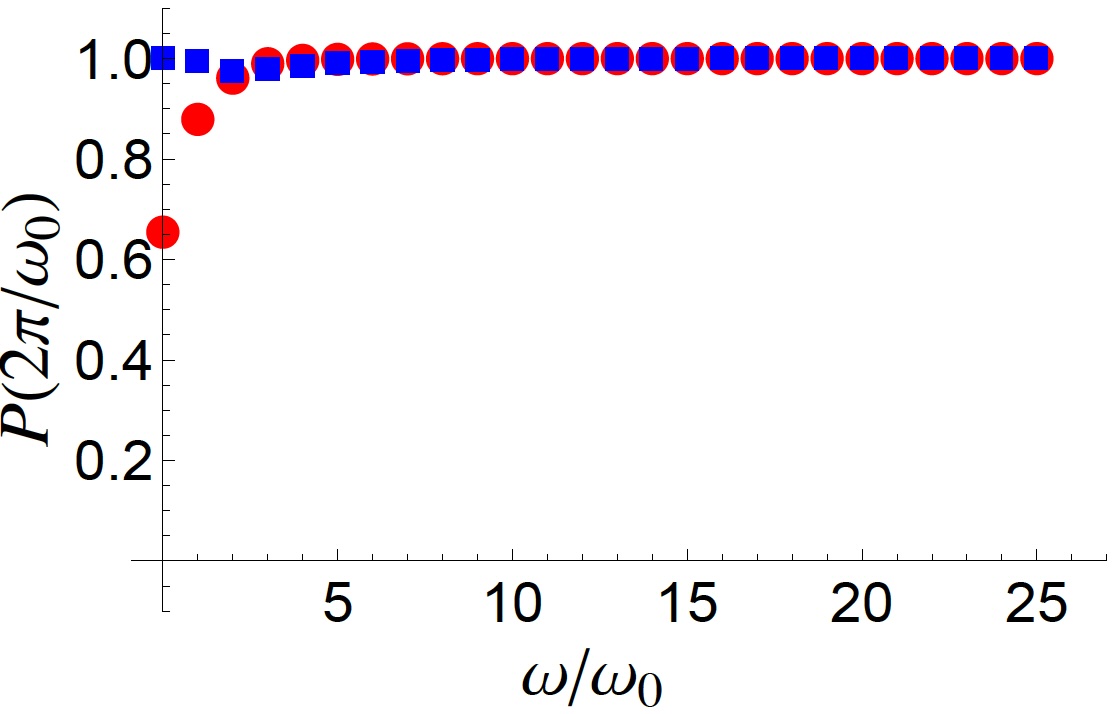}} 
\end{tabular}
\caption{Exact survival probability ($P(t)$) at $t=2\pi/\omega_0$ as a function of $k=\omega/\omega_0$, for $\lambda/\omega_0=1$, initial states $\Ket{\psi(0)}=\Ket{+}$ (red circles) and $\Ket{\psi(0)}=(\Ket{+}+\ii\Ket{-})/\sqrt{2}$ (blue squares), and for different values of $\eta$:  $\eta=0.1$ (a), $\eta=0.5$ (b), $\eta=0.9$ (c). The Hamiltonian is that in \eqref{eq:Hermitian-1}
}
\label{fig:SurvProb-Hermitian-1}
\end{figure}

\begin{figure}[h]
\centering
\begin{tabular}{ccc}
\subfigure[]{\includegraphics[width=0.30\textwidth, angle=0]{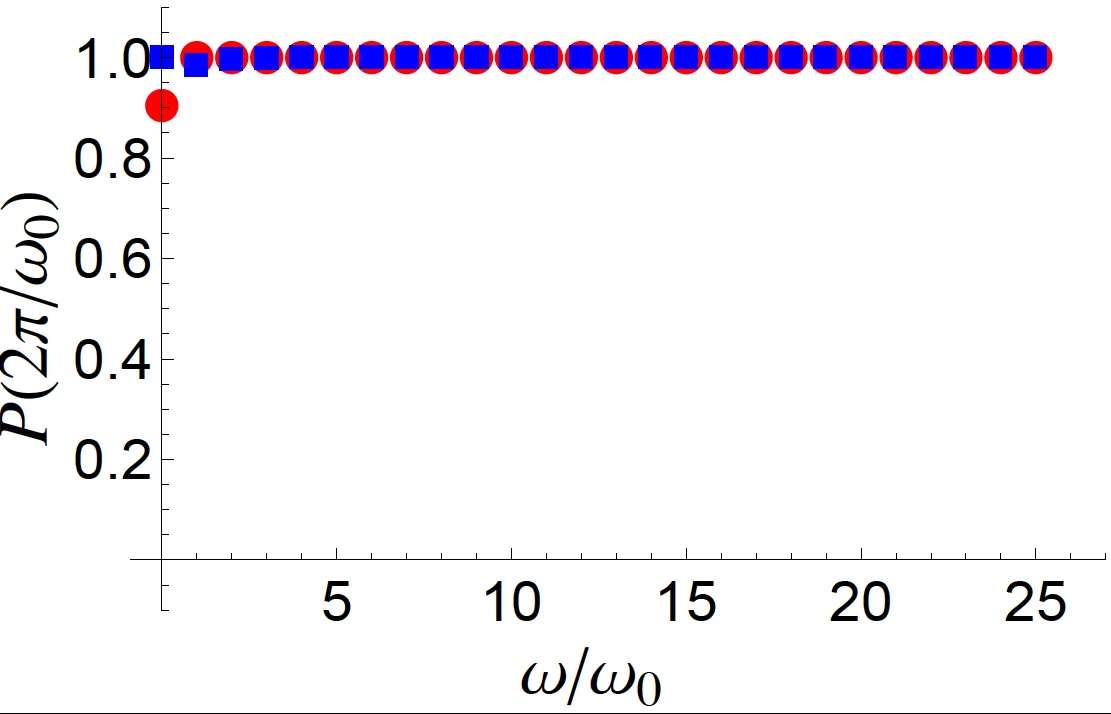}} &
\subfigure[]{\includegraphics[width=0.30\textwidth, angle=0]{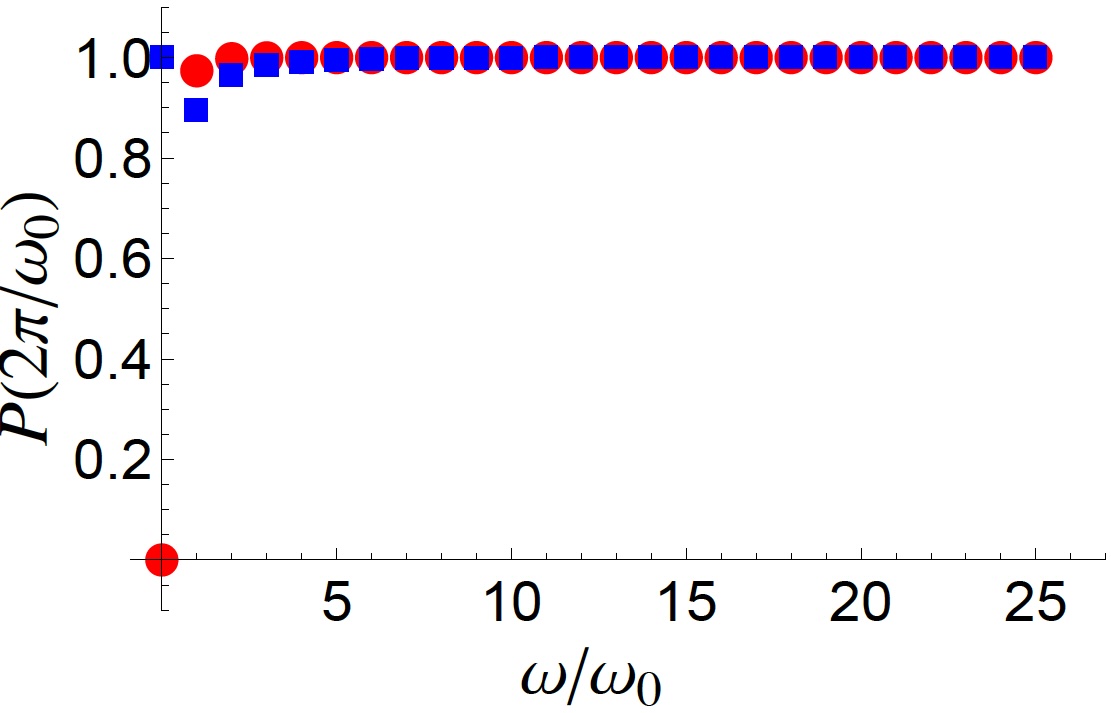}} &
\subfigure[]{\includegraphics[width=0.30\textwidth, angle=0]{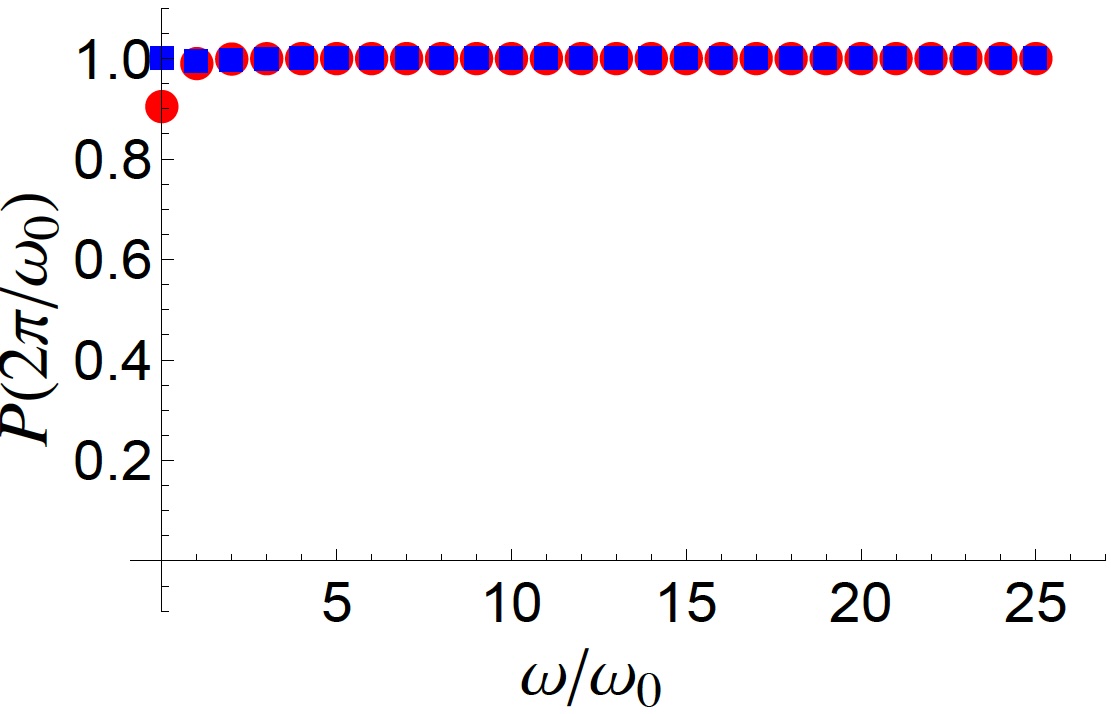}} 
\end{tabular}
\caption{Exact survival probability ($P(t)$) at $t=2\pi/\omega_0$ as a function of $k=\omega/\omega_0$, for $\lambda/\omega_0=0.5$, initial states $\Ket{\psi(0)}=\Ket{+}$ (red circles) and $\Ket{\psi(0)}=(\Ket{+}+\ii\Ket{-})/\sqrt{2}$ (blue squares), and for different values of $\eta$:  $\eta=0.1$ (a), $\eta=0.5$ (b), $\eta=0.9$ (c). The Hamiltonian is that in \eqref{eq:Hermitian-1}
}
\label{fig:SurvProb-Hermitian-2}
\end{figure}

In order to check our predictions, we consider the (numerically) exact survival probability at the time we focus on.
According to our analysis, for a given $\omega$, at time $t=2\pi/\omega$ (or any multiple of this time) the survival probability should be equal to $1$. However, with a growing $\omega$ one would have to focus on a diminishing time interval, which could trivialize our result. Therefore, we decide to fix an instant of time, $t=2\pi/\omega_0$ and, consequently, focus on frequencies such that after such a time interval the integral of the Hamiltonian vanishes, which is $\omega = k \omega_0$, with $k$ an integer number.
In Figs.~\ref{fig:SurvProb-Hermitian-1} and \ref{fig:SurvProb-Hermitian-2} it is shown the survival probability after a time $t = 2\pi / \omega_0$ as a function  $\omega/\omega_0$, for different values of the parameters and for different initial states. In all such cases, it is well visible that for high values of $\omega$ the survival probability approaches unity. On the other hand, we analytically predict a second-order survival probability equal to one for every $\omega=k\omega_0$. The discrepancy for not-so-high values of $\omega$ is traceable back to the lack of validity fo the second-order perturbation treatment our prediction is based on. 

Finally, we comment on the fact that repeatedly measuring the system at such instants of time where the integral of the Hamiltonian is zero would provide certainty of finding the system in its initial state, which resembles a perfect quantum Zeno effect. However, in our case, the time interval cannot be shortened (as required in the standard Zeno effect) because this would compromise the vanishing of the Hamiltonian integral.

{\it Non-Hermitian case: distributing or suppressing decays ---}  Let us now consider a two-state system undergoing a natural decay and subject to time-dependent fields, such that the relevant Hamiltonian can be written as follows: 
\begin{eqnarray}\label{eq:Non-Hermitian-1}
H(t) =  \frac{\omega(t)}{2} \sigma_x - \frac{\ii \Gamma }{2}(\sigma_z+\idop) + \kappa \sigma_y 
\,,
\end{eqnarray}
where $\sigma_k$'s are the standard Pauli matrices.
This kind of Hamiltonian is useful for example in describing a superconducting qubit with a state ($\Ket{+}$) undergoing a decay\cite{ref:Zhang2021}.
Assuming that the parameters $\Gamma$ and $\kappa$ are small enough, one is legitimated to estimate the survival probability with a second order perturbation theory.
In the interaction picture associated to the term proportional to $\sigma_x$ we get:
\begin{eqnarray}\label{eq:Non-Hermitian-1-2}
\tilde{H}(t) =  - \frac{\ii \Gamma }{2}( \tau_+ \ee^{\ii\Omega(t)} + \tau_- \ee^{-\ii\Omega(t)} ) 
- \ii \frac{\Gamma }{2} \idop\, + \kappa \, \ii ( \tau_+ \ee^{\ii\Omega(t)} - \tau_- \ee^{-\ii\Omega(t)} ) \,, 
\end{eqnarray}
where $\tau_\pm = \Ket{\pm}_{xx}\Bra{\mp}$ are the jump operators associated to the eigenstates of $\sigma_x$, $\Ket{\pm}_x=(\Ket{+}\pm\Ket{-})/\sqrt{2}$, and $\Omega(t) = \int_{t_1}^t \omega(s)\diff s$. 

First, one should note that the cross terms sum up to zero, which can be proven straightforwardly. 
Second, if an instant of time exists such that $\int_{t_1}^t \ee^{\ii\Omega(s)}\diff s = 0$, then it turns out that the integrals of the first and third terms of \eqref{eq:Non-Hermitian-1-2} are equal to zero, and, accordingly, the two states of the system have \lq\lq seemingly\rq\rq\, undergone a decay with a rate $\Gamma/2$, as expressed by the residual term $-\ii(\Gamma/2)\idop$, whose integral is $-\ii(\Gamma t /2)\idop$. Therefore, in this picture, the survival probability does involve only the first and second order terms associated to $\int_{t_1}^t \tilde{H}_-(s)\diff s  = \int_{t_1}^t \tilde{H}(s)\diff s  = -\ii(\Gamma t /2)\idop$. As a consequence, looking at the system at this instant of time makes it effectively evolves as if both the states are undergoing a decay with rate $\Gamma/2$, instead of having a state decaying with rate $\Gamma$ and a non-decaying state. As a technical remark, when  $\int_{t_1}^t \ee^{\ii\Omega(s)\diff s} = 0$ the unitary operator for the passage to the interaction picture is equal to the identity operator, which means that coming back to the Schr\"odinger picture in order to evaluate the survival probability will leave the state unchanged.

\begin{figure}[h]
\centering
\begin{tabular}{ccc}
\subfigure[]{\includegraphics[width=0.30\textwidth, angle=0]{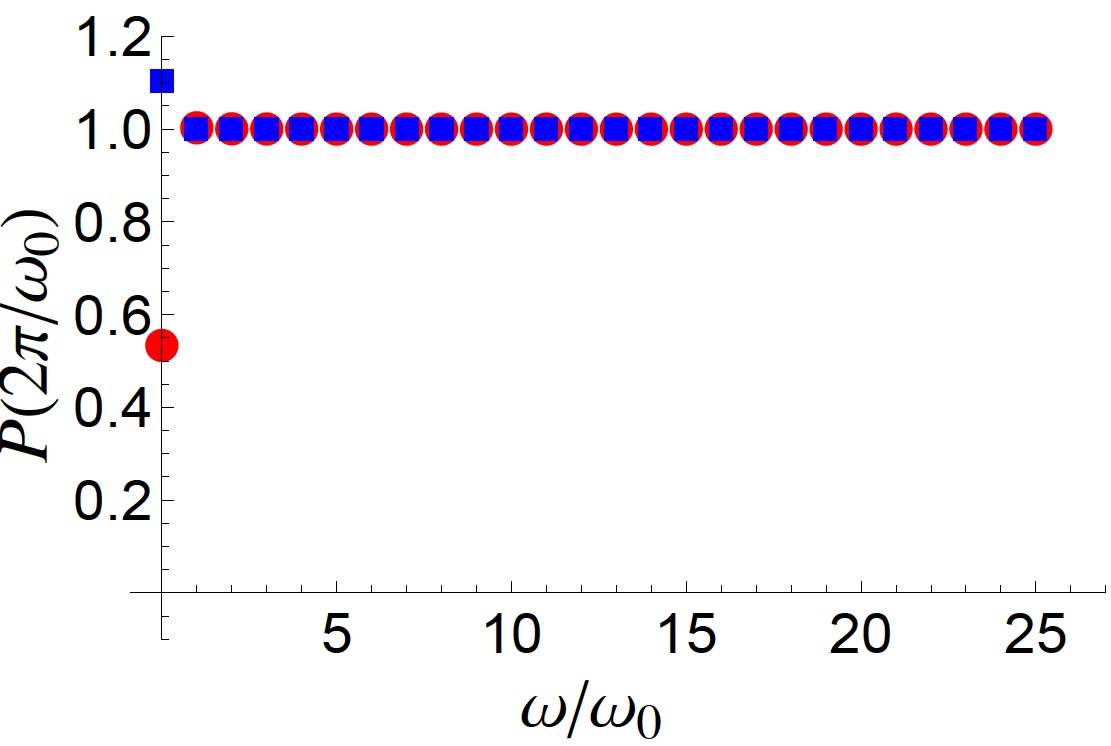}} &
\subfigure[]{\includegraphics[width=0.30\textwidth, angle=0]{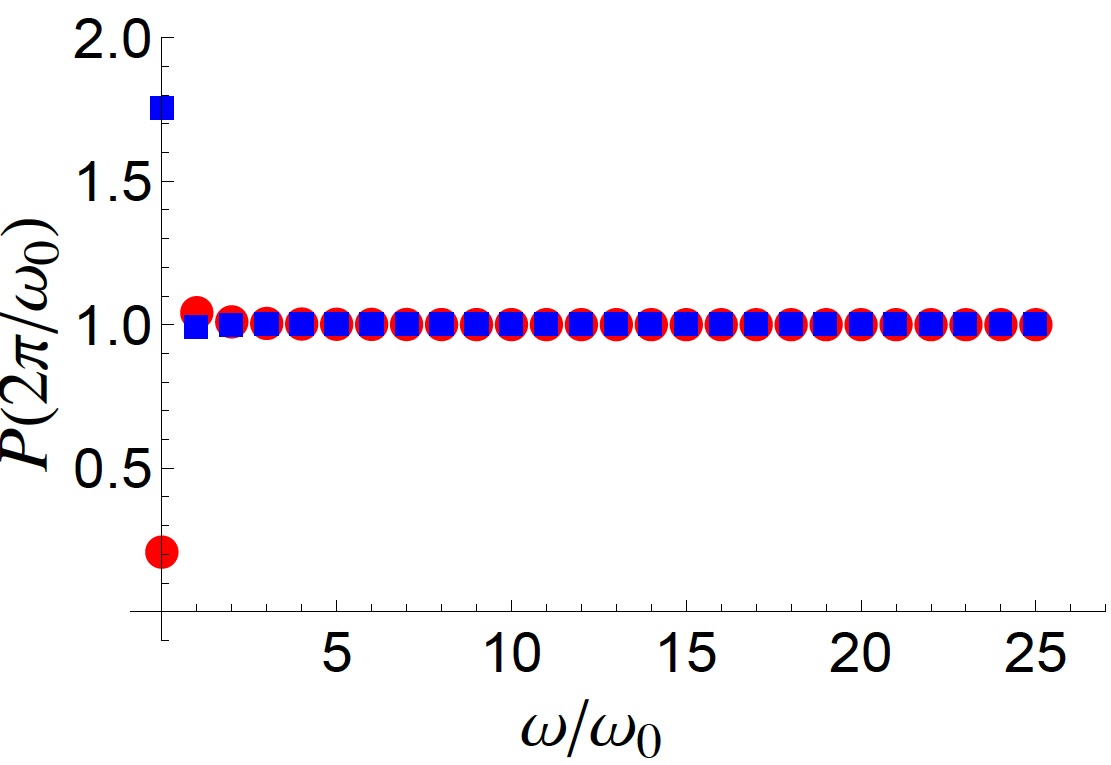}} &
\subfigure[]{\includegraphics[width=0.30\textwidth, angle=0]{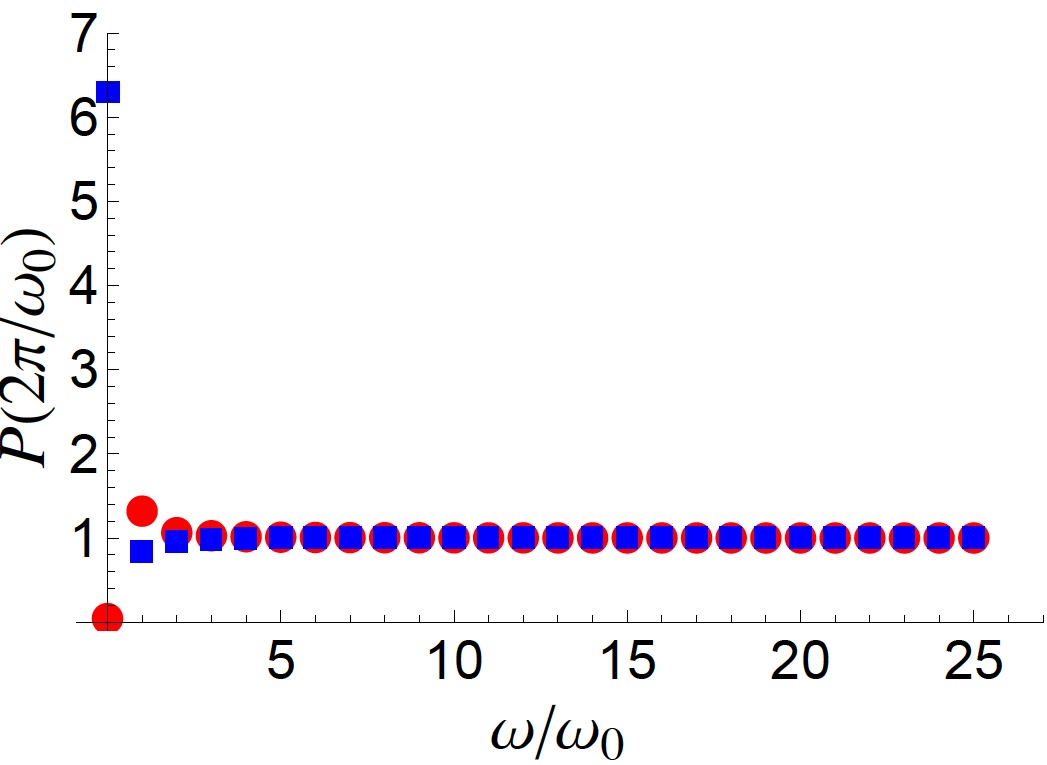}} 
\end{tabular}
\caption{Exact survival probability ($P(t)$) at $t=2\pi/\omega_0$ as a function of $k=\omega/\omega_0$, for initial states $\Ket{\psi(0)}=\Ket{+}$ (red circles) and $\Ket{\psi(0)}=(\Ket{+}+\ii\Ket{-})/\sqrt{2}$ (blue squares), for $\kappa=0$ and different values of $\Gamma$:  $\Gamma/\omega_0=0.1$ (a), $\Gamma/\omega_0=0.25$ (b), $\Gamma/\omega_0=0.5$ (c). The Hamiltonian is that in \eqref{eq:Non-Hermitian-2}
}
\label{fig:SurvProb-Non-Hermitian-1}
\end{figure}

\begin{figure}[h]
\centering
\begin{tabular}{ccc}
\subfigure[]{\includegraphics[width=0.30\textwidth, angle=0]{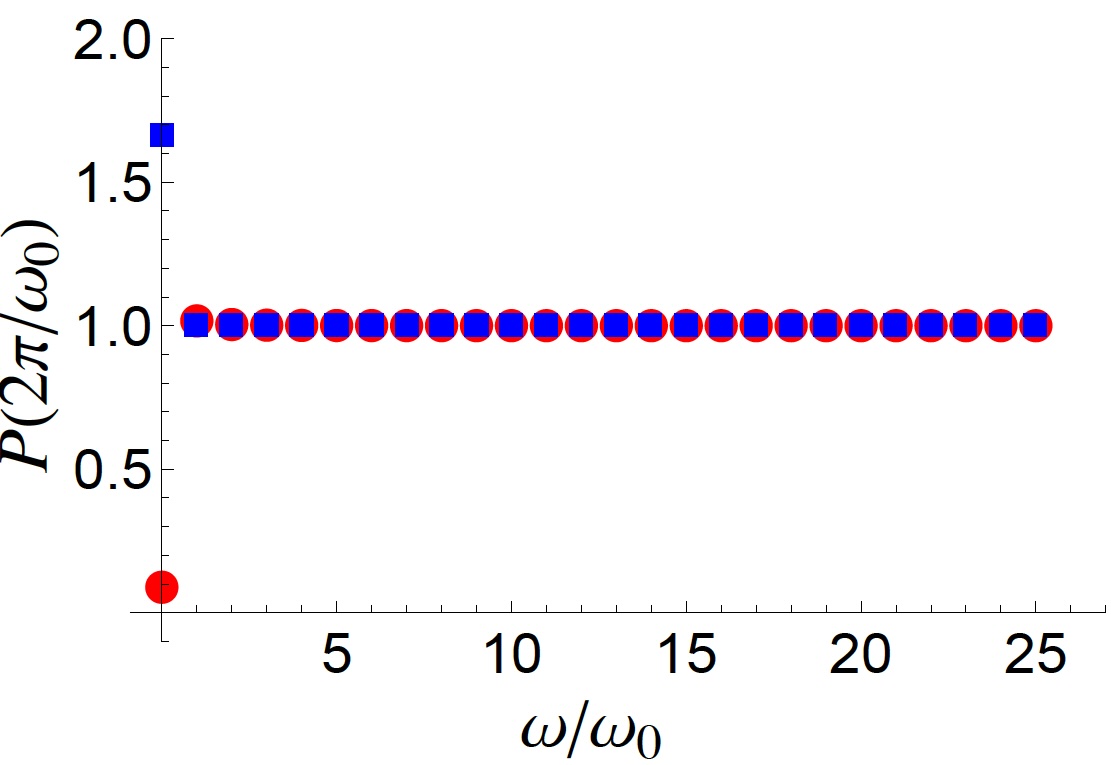}} &
\subfigure[]{\includegraphics[width=0.30\textwidth, angle=0]{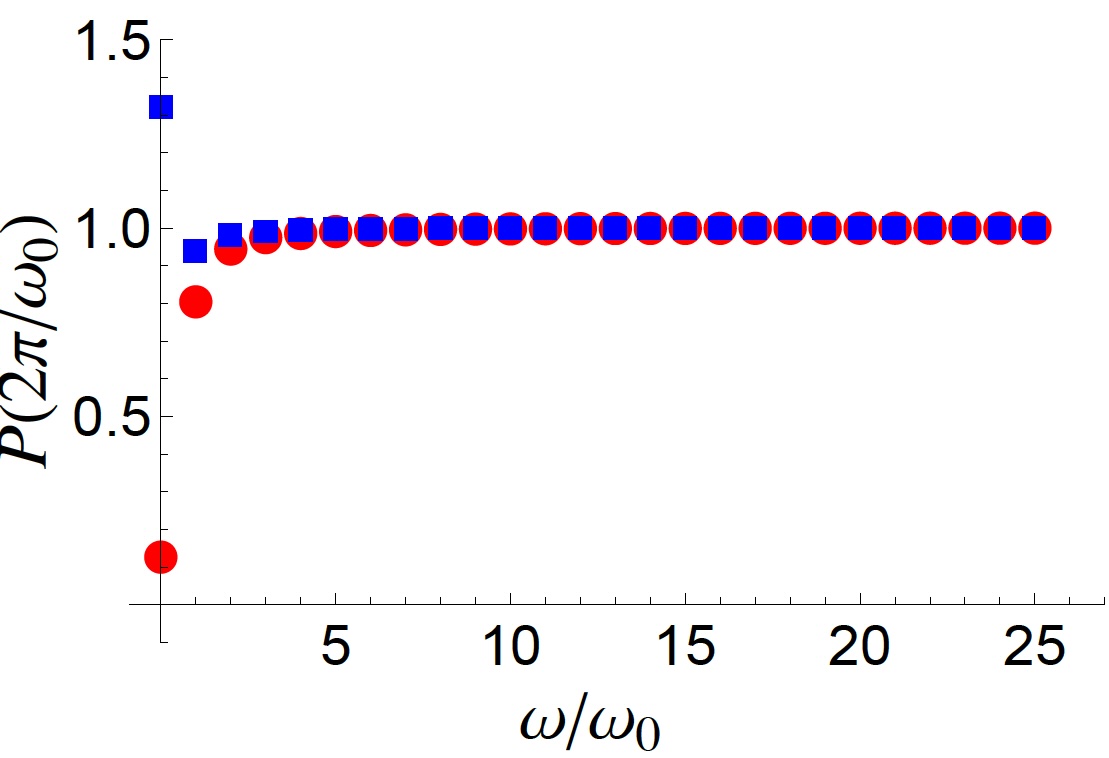}} &
\subfigure[]{\includegraphics[width=0.30\textwidth, angle=0]{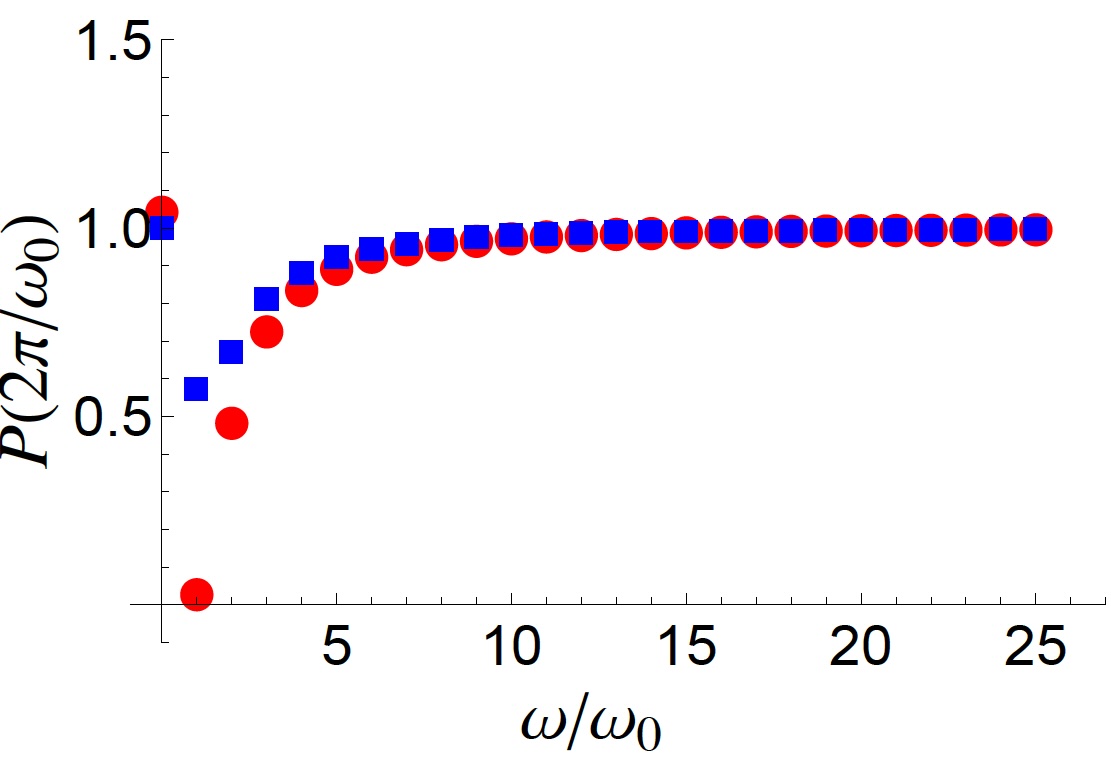}} 
\end{tabular}
\caption{Exact survival probability ($P(t)$) at $t=2\pi/\omega_0$ as a function of $k=\omega/\omega_0$, for initial states $\Ket{\psi(0)}=\Ket{+}$ (red circles) and $\Ket{\psi(0)}=(\Ket{+}+\ii\Ket{-})/\sqrt{2}$ (blue squares), for $\Gamma/\omega_0=0.25$ and different values of $\kappa$:  $\kappa/\omega_0=-0.1$ (a), $\kappa/\omega_0=-0.25$ (b), $\kappa/\omega_0=-0.5$ (c). The Hamiltonian is that in \eqref{eq:Non-Hermitian-2}
}
\label{fig:SurvProb-Non-Hermitian-2}
\end{figure}

It is worth mentioning that if the identity operator were absent from the Hamiltonian, as in the following,
\begin{eqnarray}\label{eq:Non-Hermitian-2}
H(t) =  \frac{\omega(t)}{2} \sigma_x - \frac{\ii \Gamma }{2} \, \sigma_z + \kappa \sigma_y  \,,
\end{eqnarray}
then the decay would have resulted as suppressed, at the specific time instant we have considered. Indeed, the term proportional to $\idop$ is the only surviving to the integration. However, this Hamiltonian is unphysical, involving a $+\ii\Gamma/2$ diagonal term, which would imply a general probability increase, even beyond the value $1$. Nevertheless, there are such physical systems where some average quantities satisfy a linear equation which can be considered a pseudo-Schr\"odinger equation describing both \lq\lq loss and gain\rq\rq~\cite{ref:ElGanainy2018}. 
In this situation imaginary parts of both signs are allowed for the diagonal terms of the \lq\lq pseudo-Hamiltonian\rq\rq. 
In Fig.~\ref{fig:SurvProb-Non-Hermitian-1} and ~\ref{fig:SurvProb-Non-Hermitian-2} it is shown the (numerically) exact survival probability for different initial conditions and different values of the decay rate $\Gamma$ and the coupling constant $\kappa$. In particular, Fig.~\ref{fig:SurvProb-Non-Hermitian-1} considers $\kappa=0$ and different values of $\Gamma$, while Fig.~\ref{fig:SurvProb-Non-Hermitian-2} considers a fixed value of $\Gamma$ and different values of non vanishing $\kappa$. In all these cases, for large (even not too large) values of $\omega$ we obtain that the survival probability approaches unity, since the contributions coming from the oscillating non-Hermitian part are zero, in a second order perturbation theory. By the way, it is the case to point out that a probability exceeding the value $1$ for lower values of $\omega/\omega_0$ is consistent with the presence of the \lq\lq gain term\rq\rq\, $+\ii\Gamma/2$.
 
As another example, consider the case of a static Hermitian part and an oscillating non-Hermitian part, as in the model of Ref.~\cite{ref:Lee2015}:
\begin{eqnarray}\label{eq:Non-Hermitian-3}
H = \frac{\omega_0}{2} \sigma_z - \ii \frac{\Gamma}{2} \cos(\omega t) \, \sigma_x \,, 
\end{eqnarray}
which again contains some gain mechanism, since the operator $\cos(\omega t)\sigma_x$ has time-dependent eigenvalues of opposite signs.

On the basis of this Hamiltonian, in the Schr\"odinger picture, assuming $t_1=0$, we can calculate the second-order survival as:
\begin{eqnarray}\label{eq:SP-NH-3}
\nonumber
&& |\Bra{\psi} {\cal T}(t, 0) \Ket{\psi} |^2_{(2)}  = 
1 - \Gamma \sin(\omega t) \Bra{\psi}\sigma_x\Ket{\psi} - \variance_+^2 \left( -\ii \frac{\Gamma}{2\omega} \sin(\omega t) \sigma_x, \Ket{\psi} \right) \\
&& \qquad - \variance_-^2 \left(  \frac{\omega_0 t}{2} \sigma_z , \Ket{\psi} \right)  
-  \frac{\Gamma \omega_0}{4} \Bra{\psi} \sigma_y \Ket{\psi}  \left[ \frac{2 (\cos(\omega t) - 1)}{\omega^2} - \frac{t \sin(\omega t)}{\omega}  \right] \,,
\end{eqnarray}
where the last term is the result of the double integral of the difference of the two commutators, whose temporal dependence is given by the double integral of $\int_0^t \diff s \int_0^s \diff s' (\cos(\omega s) - \cos(\omega s'))$.
In view of possible applications in the context of the quantum Zeno effect, it is worth noting that if one considers $t=2\pi/\omega$ (or an integer multiple), then the survival probability is essentially that given by the Hermitian part of the Hamiltonian, regardless of the initial state. In fact, the decay and all other non-Hermitian processes are effectively suppressed if one measures the population of the initial state at this instant of time.  
Let us now focus on a specific initial state which could be any of the two eigenstates of $\sigma_y$, and evaluate the survival probability on a generic instant of time: 
\begin{eqnarray}\label{eq:SP-NH-3}
|_y\Bra{\pm} {\cal T}(t, 0) \Ket{\pm}_y |^2_{(2)}  = 
1 + \frac{\Gamma^2}{4\omega^2}\sin^2(\omega t) - \left(  \frac{\omega_0^2 t^2}{4} \right)
\mp  \frac{\Gamma \omega_0}{4}  \left[ \frac{2 (\cos(\omega t) - 1)}{\omega^2} - \frac{t \sin(\omega t)}{\omega}  \right] \,.
\end{eqnarray}
It is noteworthy that no first order decay term appears in this expression.


\section{Discussion}\label{sec:Discussion}

In this paper we have presented a formula for the second-order survival probability which allows evaluating short-time behaviors of quantum systems which are not necessarily closed and isolated. The focus on second-order is due to the fact that the standard treatment of the quantum Zeno effect is based on a second-order perturbative approach, which we have extended to cases that go beyond static Hermitian Hamiltonians. Indeed, our formula applies to every kind of non-Hermitian Hamiltonian (with either real or complex eigenvalues), even in the case where they are time-dependent, and involves the variance and pseudo-variance of the integrals of the Hermitian and anti-Hermitian parts of the Hamiltonian, respectively. In addition, it also involves a difference between two commutators of the Hermitian and anti-Hermitian parts of the Hamiltonian at different times, which somehow encodes the interplay between unitary and non-unitary processes.

In the case of Hermitian time-dependent Hamiltonians we are able to predict a sort of Zeno effect, which, unlike the standard version of the QZE and possible generalization to time-dependent Hamiltonians, does not necessarily require the time intervals between measurements to go to zero. On the contrary, we usually require a finite time. In fact, we are able to identify, under suitable hypotheses, instants of time where the survival probability is equal to unity, even if it has assumed lower values at earlier instants of time. This is easily achieved by considering that deviations of the second order survival probability from unity are given by the variance of the integral of the Hamiltonian in the initial state. Therefore, once such an integral vanishes, the approximate survival probability is equal to one, regardless of the initial state.

In the presence of a non-Hermitian Hamiltonian, an effective alteration of the decay (or other non-unitary) processes can be obtained. In some cases, a decay which is relevant to a single state can be somehow \lq\lq distributed\rq\rq\, over all the other states: this is what happens to a two-state system with a decaying state and a stable one, which exhibits an overall decay with half the decay rate of the unstable state. In some other cases, all the non-unitary processes can be suppressed. This is the case of a non-Hermitian term describing gain and loss at equal rates.

\begin{figure}[h]
\centering
\begin{tabular}{ccc}
\subfigure[]{\includegraphics[width=0.33\textwidth, angle=0]{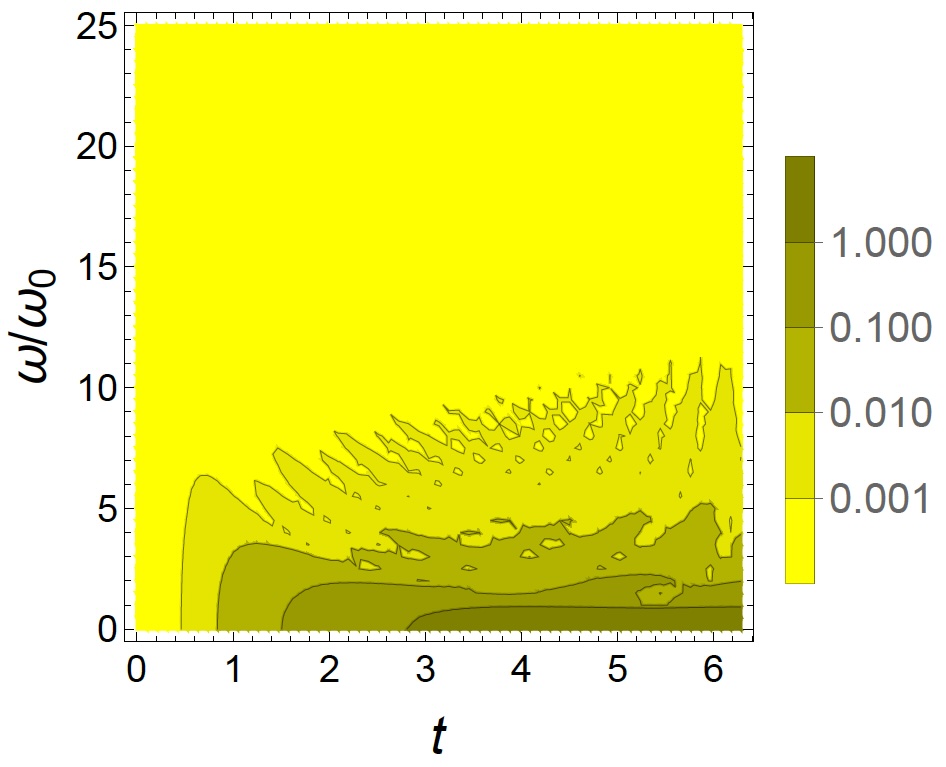}} &
\subfigure[]{\includegraphics[width=0.33\textwidth, angle=0]{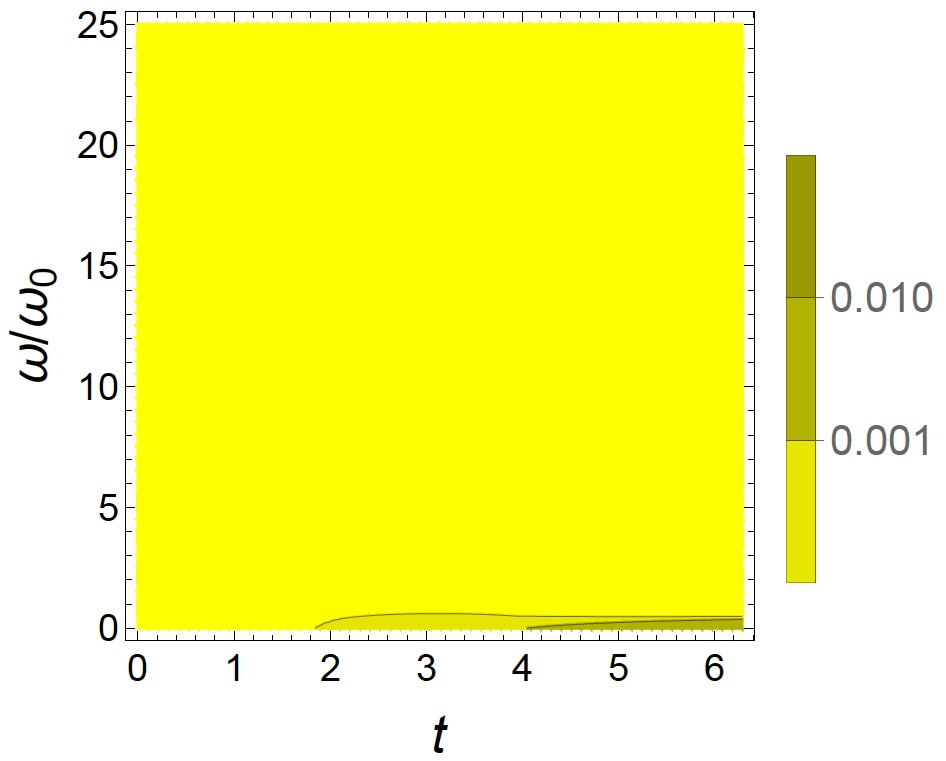}} &
\subfigure[]{\includegraphics[width=0.33\textwidth, angle=0]{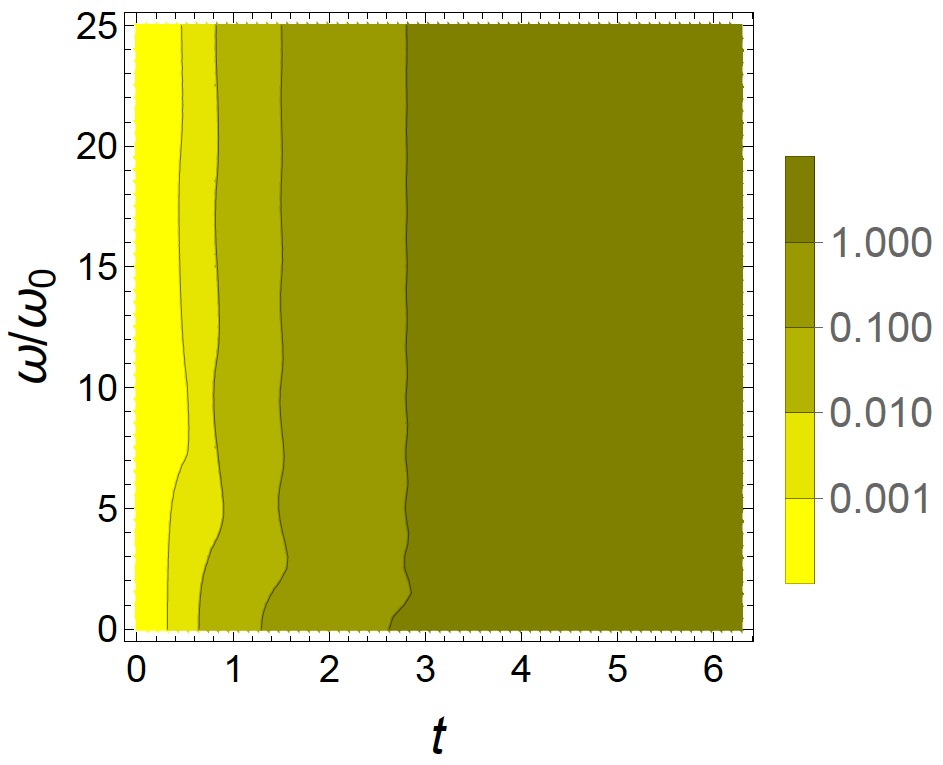}} 
\end{tabular}
\caption{The absolute value of the difference between the (numerically) exact survival probability and the survival probability evaluated via the second order perturbation, $Abs(|\Bra{\psi} {\cal T}(t_2, t_1) \Ket{\psi} |^2 - |\Bra{\psi} {\cal T}(t_2, t_1) \Ket{\psi} |^2_{(2)})$, as a function of the ratio $\omega/\omega_0$ and time $t$ in units of $1/\omega_0$. This quantity is plotted for different models and initial states. 
In (a) the model is the one in \eqref{eq:Hermitian-1}, with $\lambda/\omega_0=1$, $\eta/\omega_0=0.5$ and initial state $\Ket{\psi(0)}=\Ket{+}$.
In (b) and (c) the model is that in \eqref{eq:Non-Hermitian-3} with $\Gamma/\omega_0 = 0.1$ and initial state $\Ket{\psi(0)}=\Ket{+}$ in (b) and $\Ket{\psi(0)}=(\Ket{+}+\Ket{-})/\sqrt{2}$ in (c).
}
\label{fig:Discrepancy}
\end{figure}
An important comment concerns the validity of our approximation. Indeed, as we have pointed out several times, our treatment is based on a perturbative approach and hence is valid under suitable hypotheses involving relevant parameters. This issue clearly emerges from the plots presented here, where a survival probability is predicted to be equal to one even for small values of the frequency $\omega$, while the numerical resolution of the dynamics (reported in the plots) shows a potentially significant discrepancy, which rapidly decreases for higher values of the frequency $\omega$. Indeed, in the examples analyzed here, a higher $\omega$ ensures the validity of the second-order perturbative treatment.
As an example consider the Hermitian model in \eqref{eq:Hermitian-1}, where the whole Hamiltonian is oscillating at frequency $\omega$. As usual, the perturbative character of the oscillating terms is established once the amplitude is much smaller than the frequency, and, since $\lambda=\omega_0$, this corresponds to having $\omega\gg \omega_0$.
A similar comment can be made concerning the non-Hermitian time-independent models in \eqref{eq:Non-Hermitian-1} and \eqref{eq:Non-Hermitian-2}, once they are analyzed in the interaction picture, where they turn out to oscillate at frequency $\omega$. Therefore, $\kappa$ and $\Gamma$, which are of the order of $\omega_0$, are supposed to be much smaller than $\omega$.
The non-Hermitian time-dependent model in \eqref{eq:Non-Hermitian-3} deserves a slightly more complicated discussion. Indeed, while the oscillating part of the Hamiltonian can be treated as a perturbation, according to previous considerations, the stationary part cannot. Nevertheless, if we consider small intervals of time, a second-order expansion of the unitary evolution operator associated to the sole stationary part would be allowed. Therefore, if we mainly focus on time $t=2\pi/\omega$, for large $\omega$ we can satisfy both the conditions (smallness of perturbation, second-order treatment for the larger part).
In order to strengthen these assertions about the validity of our approximation, we have made some plots comparing the numerically exact evaluation of the survival probability, with the same quantity calculated on the basis of our formula obtained via a perturbative treatment, beyond the specific instants where perfect revival of the initial state is predicted. In Fig.~\ref{fig:Discrepancy} the absolute value of the difference $|\Bra{\psi} {\cal T}(t_2, t_1) \Ket{\psi} |^2 - |\Bra{\psi} {\cal T}(t_2, t_1) \Ket{\psi} |^2_{(2)}$ is shown, where ${\cal T}(t_2, t_1)$ is the exact evolution operator, here numerically evaluated. In particular, in Fig.~\ref{fig:Discrepancy}a we focus on the model in \eqref{eq:Hermitian-1}, showing a very good agreement between the exact and approximate survival probability (discrepancy $<0.001$), especially for larger values of $\omega$. In  Figs.~\ref{fig:Discrepancy}b and \ref{fig:Discrepancy}c  we focus on the model in \eqref{eq:Non-Hermitian-3} and it clearly emerges that in this case the validity of our treatment is state-dependent. This circumstance is related to the fact that we have a non perturbative term ($\omega_0/2 \, \sigma_z$), which influences the dynamics in different ways, depending on whether the initial state is an eigenstate of such operator ($\Ket{+}$,  in Fig.~\ref{fig:Discrepancy}b) or not ($(\Ket{+}+\Ket{-})/\sqrt{2}$,  in Fig.~\ref{fig:Discrepancy}c). Nevertheless, it is important to stress that, regardless of the state, for small $t$, and preferably large $\omega$,  the agreement is very good, according to the previous comments.

\end{document}